\documentstyle[epsfig,osa,twocolumn]{revtex}

\begin{document}
%preprint{draft 2.3.00}
\preprint{submitted to Opt. Lett. 23.3.00}

\title{Frequency stabilization of a monolithic Nd:YAG ring laser by controlling
the power of the laser-diode pump source}

\author{B.\,Willke , S.\,Brozek, and K.\,Danzmann}
\address{Institut f\"ur Atom- und Molek\"ulphysik, Universit\"at Hannover,
Callinstr. 38, D--30167 Hannover % , Germany \\
\\ Max-Planck-Institut f\"ur Quantenoptik, Hans-Kopfermann-Str.\,1,
D--85748 Garching, Germany} 
%$^3$Laser Zenrum Hannover e.V., Hollerithalle 8, D-30419 Hannover, Germany}

\author{V.\,Quetschke, and S.\,Gossler }
\address{Institut
    f\"ur Atom- und Molek\"ulphysik, Universit\"at Hannover,
    Callinstr. 38, D--30167 Hannover , Germany\\ [2mm] 
\tt accepted for publication in Opt.\ Lett. April 26, 2000 }

%$^3$Laser Zenrum Hannover e.V., Hollerithalle 8, D-30419 Hannover, Germany}

\maketitle

%\bf Korrelationen zwischen Frequenz und Leistungsfluktuationen messen

\begin{abstract}
The frequency of a 700\,mW monolithic non-planar Nd:YAG ring laser (NPRO)
depends with a large coupling coefficient (some $\mbox{MHz}/ \mbox{mW}$) on the
power of its laser-diode pump source. Using this effect we demonstrate the
frequency stabilization of an NPRO to a frequency reference by feeding back to
the current of its pump diodes.

We achieved an error point frequency noise smaller than $ 1\,\mbox{mHz}
/ \sqrt{\mbox{Hz}} $, and simultaneously a reduction of the power noise of the
NPRO by 10\,dB without an additional power stabilization feed-back system.  

\end{abstract}

\section*{}
Due to the demanding requirements of recent experiments in quantum optics,
laser spectroscopy and laser metrology there has been much interest in laser
% stabilization over the last years. Although very good stability results were
stabilization over the last years. Although very good stability was
achieved with $ \mbox{Ar}^+ $ lasers~\cite{Hough91}, diode-laser pumped
solid-state lasers were chosen for almost all modern high-precision experiments.
The reason for this choice is  that the free-running frequency noise of
these lasers is 2 to 3 orders of magnitude smaller than for $\rm Ar^+$
lasers, and the intensity noise of solid-states-laser is also much
lower. Furthermore, solid-state lasers have a very high electrical-to-optical
efficiency, which is important especially in space applications like 
inter-satellite communication or high power applications as laser 
interferometric gravitational wave detectors.

Many of these experiments rely on the high intrinsic stability of Nd:YAG
non-planar ring oscillators (NPRO)\cite{Kane85}, the output of which is used in
the experiment directly or is amplified either by injection locking
\cite{Farinas95,Golla96,Ottaway98} 
%% or in a master-oscillator power-amplifier configuration \cite{Willke98}. 
 or in a configuration with master oscillator and power amplifier
\cite{Willke98}. 
The free-running frequency noise spectral
density of NPROs is of the order of $\mbox{1 kHz} / \sqrt{\mbox{Hz}}$ at 10 Hz
and falls like $1 / f$  at higher frequencies. The unstabilized power
noise of such lasers has a level of $ 10^{-7} / {\sqrt{\mbox{Hz}}} $. Although
this intrinsic stability is quite high, experiments like gravitational wave
% detectors require a frequency stability which is in the $\mbox{mHz} /
detectors require a frequency stability in the $\mbox{mHz} /
\sqrt{\mbox{Hz}}$ range, and simultaneously the power noise needs to be reduced
by at least an order of magnitude.

The commonly used schemes to reduce the frequency noise of NPROs rely on 
stabilizing the laser frequency to a fixed-spacer reference cavity or an atomic
resonance by feeding back to two different actuators: the temperature of the
Nd:YAG crystal in the low Fourier frequency range below 1\,Hz, and for higher
frequencies to a piezo-electrical transducer (PZT) mounted on top of the
crystal that changes the laser frequency due to stress-induced
birefringence. The resonances of the PZT above 100\,kHz limit the useful
bandwidth of the latter actuator. Good results were achieved especially by using
an additional external phase shifter (Pockels cell) as a fast actuator to
increase the unity gain frequency of the feed-back control loop up to
1\,MHz. For example Bondu et al. \cite{Bondu96} report a frequency noise
spectral density below $10^{-4}\, \mbox{Hz} / {\sqrt{\mbox{Hz}}} $ with respect
to the reference cavity (in-loop) and in the order of $10^{-2}\, \mbox{Hz} /
{\sqrt{\mbox{Hz}}} $ with respect to an independent cavity
(out-of-loop). 

Although these results already meet the demanding requirements of first
generation gravitational wave detectors, no attention was paid to the power
noise and spatial beam fluctuations. Currently performed cross coupling
measurements \cite{Quetschke00} predict a non-negligible pointing and also
power noise introduced by feeding a signal to the NPRO's PZT. Furthermore,
care has to be taken that residual amplitude modulation of the
phase-correcting Pockels cell does not compromize the shot noise limited
performance of the NPRO in the frequency range above 5\,MHz, which is essential
for the heterodyne detection scheme used in many experiments.  On the other
hand the power stabilization scheme normally employed adds a signal to the
current of the pump source of the NPRO, which has the undesired effect of
changing the NPRO frequency. These problems together with the understanding of
the fact that the free-running frequency noise of the NPRO is mainly due to
power fluctuations of the laser-diode pump source \cite{Day90} led us towards
the new stabilization scheme. (A related scheme with a separate heating laser was used by Heilmann et
al. \cite{Heilmann92} to stabilize a twisted-mode-cavity laser.)

\begin{figure} % fig 1
\centering \epsfig{file=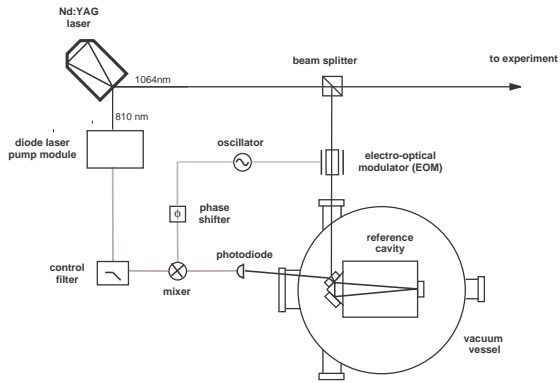,width=100mm,clip=}% \hspace{10cm}
\caption{\label{fig1}Schematic of the experimental setup.  An NPRO is stabilized to a
  rigid-spacer reference cavity by using the Pound-Drever-Hall scheme to
  achieve an error signal for the control loop feeding back to the current of
  the NPRO's pump diode.}

\end{figure}

 Figure~\ref{fig1} shows a sketch of the experimental setup.  A 700\,mW NPRO
 built by     Laser Zentrum Hannover was mode-matched to a high-finesse
 fixed-spacer ring cavity made from ultra-low-expansion material (ULE). This
 resonator has a finesse of 58\,000 and is put in a vacuum tank to avoid 
 contamination and acoustic disturbances. 

 Before entering the cavity the light passes through a  
% optical diode to avoid back-reflections of the light into the laser
 Faraday isolator to avoid back-reflections of the light into the laser
 and is transmitted through an electro-optical modulator (EOM). The EOM was
 driven by an rf-oscillator at $ \omega_{\rm mod} = 29\, \mbox{MHz} $ to produce
   %% einfacher ginge es mit:  $\omega{\rm mod} = 29\,$MHz                
 phase modulation sidebands on the light. Once the laser frequency 
 $\omega_{\rm L}$
 is near a resonance $\omega_{\rm C}$ of the cavity, an asymmetry between these
 sidebands and the reflected carrier produces an amplitude modulation of the
 light at $\omega_{\rm mod}$ which then, detected by an InGaAs photodiode and
 demodulated at $\omega_{\rm mod} $, gives an error signal for the frequency
 stabilization servo system. A phase shifter between the rf-oscillator and
 mixer is used in this well established Pound-Drever-Hall locking scheme
 \cite{Drever83} to optimize the slope of the frequency error signal.

\begin{figure}
\centering \epsfig{file=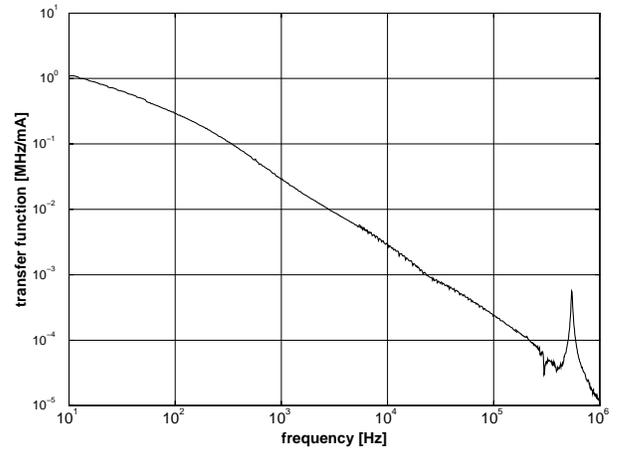,width=60mm,clip=,angle=270} 
\caption{\label{fig2}Transfer function $ T_{\rm cur \rightarrow \omega}$ 
of a signal summed to
the current of the NPRO's pump diode and the NPRO frequency. Power fluctuations
of the pump diodes together with this strong coupling are responsible for the
free-running laser frequency noise and allows the frequency stabilization of 
an NPRO by changing the power of its pump source.}

\end{figure}

To measure the transfer function $T_{\rm cur\rightarrow\omega}$ between a
signal added to the current of the pump diodes and the frequency of the NPRO we
first locked the laser to the reference cavity using the conventional method of
feeding the filtered error signal back to the PZT frequency actuator of the
NPRO. The gain of this feed-back loop was reduced to give a unity gain
frequency of only 100\,Hz. This servo was necessary to keep the laser frequency
within the central part of the cavity linewidth, as only under this condition
the Pound-Drever-Hall error signal is proportional to the difference frequency
$\delta\omega = \omega_{\rm L} -\omega_{\rm C}$ between laser and cavity
resonance.  To measure $ T_{\rm cur \rightarrow \omega}$, which is shown in
Figure~\ref{fig2}, we summed the source signal of a network analyzer with
Fourier frequency above the unity gain frequency of the servo to the
laser-diode pump-current and measured the change of the laser frequency at the
error point of the Pound-Drever-Hall circuit.

%Figure~\ref{fig2} shows a Bode diagram of this transferfunction $ T_{cur
%\rightarrow \omega}$.

%for a 700\,mW NPRO built by the Laser Zentrum Hannover.
According to a model by Day et al. \cite{Day90} this coupling is due to 
thermally induced changes of the optical path length in the laser crystal. 
% maybe next paragraph is not needed
By calculating the optical path length change of a typical NPRO crystal due to
sinusoidial power fluctuations of its pump diode, Day et al. were able to
model the transfer function $T_{\rm cur\rightarrow\omega}$ with good agreement
to the experimental result between 100\,Hz and 100\,kHz. Furthermore by
assuming a flat spectral density of the power fluctuations of the pump LDs
their model was able to predict the free-running frequency fluctuations of the
NPRO.
% down to here
Our measurements in Figure~\ref{fig2} shows $T_{\rm cur \rightarrow \omega}$ to
Fourier frequencies up to 1\,MHz, and particularly in the frequency range of
the NPRO's relaxation oscillation frequency. It is worth mentioning that the
power fluctuations of the pump LD which drive the relaxation oscillations cause
a resonant response not only in the NPRO power but also in its frequency. This
is clearly not a thermal effect but probably due to changes in the index of
refraction caused by oscillations in the atomic polarization of the active
laser medium.

\begin{figure}
\centering \epsfig{file=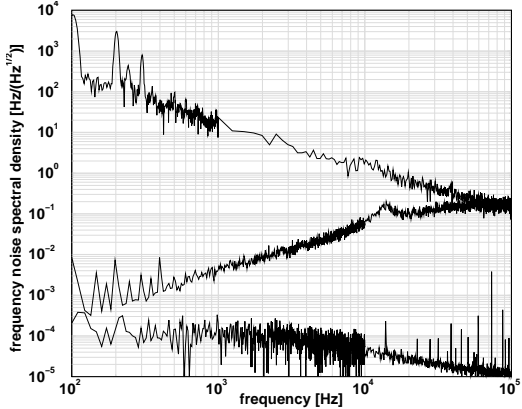,width=55mm,clip=,angle=270}% \hspace{10cm}
\caption{\label{fig3}Frequency noise spectral density of the monolithic Nd:YAG laser
  relative to a rigid-spacer reference cavity. The upper curve shows
  the free running noise and the middle curve is a measurement taken at
  the error point of a feedback loop that stabilizes the laser
  frequency by feeding back to the current driving its laser-diode
  pump source. The lower curve represents the electronic noise of the
  measurements.}

\end{figure}

Based on this transfer function we designed a control system to lock the laser
frequency $\omega _{\rm L}$ to the cavity resonance $\omega _{\rm C}$ by
feeding back to the pump LD current. Figure~\ref{fig3} shows the spectral
density of frequency fluctuations $\delta \omega = \omega_{\rm L} - \omega
_{\rm C} $.

The upper curve shows the free-running frequency noise of the NPRO
and the lower curve was measured at the error point of the closed frequency
stabilization loop.  The bandwidth of this control system was 80\,kHz and the
frequency fluctuations could be reduced to below 10$\,\rm mHz / \sqrt{Hz}$ for
Fourier frequencies below 2\,kHz, which is comparable to the noise reduction
we could achieve with a conventional split control loop feeding back to the
laser PZT and also a phase correcting Pockels cell behind the laser.

%Due to the coupling between the laser frequency and the power of the LD pump
%source the frequency measurement simultaneously measures the power fluctuations
%of the pump LD integrated over the spatial profile of the laser gain. Due to
%interferences of the different elements of the LD bar this information can not
%be achieved by a measurement of the LD pump light on a photodiode
%\cite{Harb97}. As these power fluctuations are the main source for the laser
%frequency noise the frequency stabilization servo reduces simultaneously the
%power noise of the NPRO.

\begin{figure}
\centering \epsfig{file=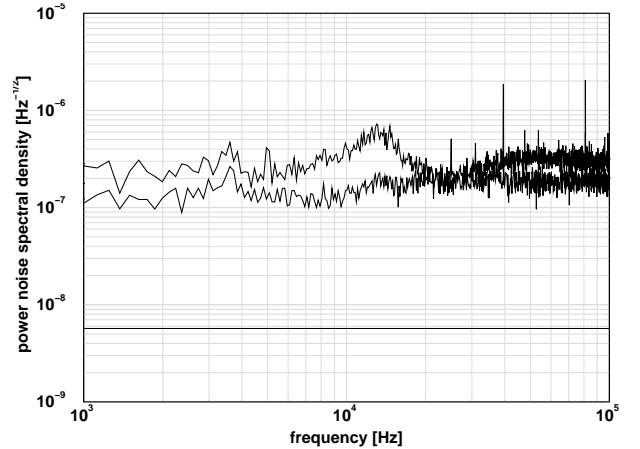,width=60mm,clip=,angle=270}% \hspace{10cm}
\caption{\label{fig4}Spectral density of the power fluctuations of a monolithic Nd:YAG
  laser. The upper curve (at low Fourier frequencies) shows the noise
  of the unstabilized laser and the middle curve was measured while the
  laser frequency was stabilized using the current lock. Both
  measurements were taken without any dedicated active intensity
  stabilization. The lower straight line corresponds to the shot noise
  limit of this measurement.}

\end{figure}

Due to the coupling between the laser frequency and the power of the LD pump
source, the frequency measurement simultaneously measures the power
fluctuations of the pump LD integrated over the spatial profile of the laser
gain.  Therefore this frequency servo simultaneously reduces the power noise of
the NPRO.  Figure~\ref{fig4} shows the power noise with and without frequency
servo closed. Although the noise is reduced significantly, there is less noise
reduction than the servo-system gain would suggest. This is probably due to the
fact that the spatial overlap between the laser volume and the pump volume is
not perfect. This means that a fraction of the absorbed pump-light can deposit
heat in the Nd:YAG crystal but does not change the gain in the laser
volume. Hence fluctuations of this pump light may cause frequency fluctuations
by changing the index of refraction of the crystal but do not change the laser
power. As the spatial distribution of the power fluctuations is not constant
over the beam profile of the laser-diode bar, the correlation between the power
and frequency fluctuations of the NPRO caused by pump-power fluctuations is not
perfect.

% Furthermore, sensing noise in the frequency servo could be the reason
% for the lacking reduction of the NPRO power fluctuations.
%for the less than expected reduction of the NPRO power fluctuations.

To summarize: we have introduced a new scheme to stabilize the frequency of an
NPRO laser which only uses the built-in laser-diode current actuator. One
advantage of this method in comparison with the conventional scheme is the much
simpler controller needed. Only low voltage feed-back electronics is used and
no cross-over has to be designed between the PZT and an external Pockels cell
actuator, both of which need notoriously noisy high-voltage
amplifiers. Furthermore, beam pointing and power fluctuations which are
introduced by the PZT could be avoided.

The detection of the frequency noise is a
non-demolition measurement of the NPRO power fluctuations and should in
principle avoid the 3\,dB penalty that is the minimum to be paid in
conventional power stabilization servos\cite{Harb97}. However,
% different
%processes than LD power noise are causing frequency noise of the NPRO and 
more investigations are needed to clarify why the power noise reduction in our
experiment was smaller than expected.

We would like to thank A. R\"udiger for his assistance in the preparation of this
manuscript. This work was supported by the Deutsche Forschungsgemeinschaft
within the Sonderforschungsbereich 407.


\begin{references}

{\frenchspacing
\bibitem{Hough91}
J.\,Hough, H.\,Ward, G.A.\,Kerr, N.L.\,Mackenzie, B.J.\,Meers,
G.P.\,Newton, D.I.\,Robertson, N.A.\,Robertson, and R.\,Schilling,
in D.G.\,Blair, ed. {\it The Detection of Gravitational Waves},
(Cambridge University Press, Cambridge, 1991)

\bibitem{Kane85}
T.\,J.\,Kane and R.\,L.\,Byer, Opt. Lett. {\bf 10}, 65 (1985)

\bibitem{Farinas95}A.\,D.\,Farinas, E.\,K.\,Gustafson, and R.\,L.\,Byer, 
J.~Opt. Soc. Am. {\bf 12}, 328, (1995)

\bibitem{Golla96}I.\,Freitag, D.\,Golla, S.\,Knoke, W.\,Sch\"one, H.\,Zellmer,
A.\,T\"unnermann, and H.\,Welling, Opt. Lett. {\bf20}, 462 (1995)

\bibitem{Ottaway98} D.\,J.\,Ottaway, P.\,J.\,Veitch, M.\,W.\,Hamilton,
C.\,Hollitt, D.\,Mudge, and J.\,Munch, IEEE J.~ Quantum Electron. {\bf 34},
2006 (1998)

\bibitem{Willke98} B.\,Willke, N.\,Uehara, E.\,K.\,Gustafson, R.\,L.\,Byer,
 P.\,J.\,King, S.\,U.\,Seel, R.\,L.\,Savage,\,Jr., Opt. Lett. {\bf 23}, 1704
  (1998)
  
\bibitem{Bondu96} F.\,Bondu, P.\,Fritschel, C.\,N.\,Man, and A.\,Brillet,
Opt. Lett. {\bf 21}, 582 (1996)

\bibitem{Quetschke00} V.\,Quetschke, in preparation

\bibitem{Day90}T.\,Day, Ph.D. thesis, Stanford University, USA (1990)

\bibitem{Heilmann92}R.\,Heilmann and B.\,Wandernoth, Electron. Lett. {\bf 28},
1367 (1992)

\bibitem{Drever83}
R.\,W.\,P.\,Drever, J.\,L.\,Hall, F.\,V.\,Kowalski, J.\,Hough,
G.\,M.\,Ford, A.\,J.\,Munley, and H.\,Ward, Appl. Phys. B {\bf 31}, 97 (1983)


\bibitem{Harb97} C.\,C.\, Harb, T.\,C.\, Ralph, E.\,H.\, Huntington, D.\,E.\,
McClelland, H.\,A.\, Bachor, and I.\,Freitag, JOSA B {\bf 14}, 2936 (1997)


%\bibitem{Skeldon96}
%K.\,D.\,Skeldon, K.\,A.\,Strain, A.\,I.\,Grant and J.\,Hough,
%Rev. Sci. Instrum. {\bf 67}, 2443 (1996)

}   %% end of \frenchspacing
\end{references}
\end{document}